\documentclass{elsart}
\usepackage{natbib}
\usepackage{epsfig}
\usepackage{subfigure}

\begin{document}

\begin{frontmatter}
\title{Three dimensional imaging of short pulses}
\author[Milano]{Marco A.C. Potenza}
\author[Barcellona]{Stefano Minardi}
\author[Como]{Jose Trull}$^{\dagger}$
\author[Como]{Gianni Blasi}
\author[Como]{Domenico Salerno}
\author[Como]{Paolo Di Trapani}
\author[Vilnius]{Arunas Varanavi\v{c}ius}
\author[Vilnius]{Algis Piskarskas}

\address[Milano]{Department of Physics "Aldo Pontremoli", University of Milan and \\
Istituto Nazionale Fisica della Materia, Via Celoria 16, 
I-20133 Milano, Italy}
\address[Barcellona]{Instituto di Ciencias Fotonicas c/Jordi Girona, 29 - 
NEXUS II E-08034 Barcelona, Spain}
\address[Como]{Istitito Nazionale Fisica della Materia, Dipartimento di 
Scienze Chimiche, Fisiche, Matematiche, Universit\`a dell'Insubria, 
Via Valleggio 11, I-22100 Como, Italy}
\address[Vilnius]{Department of Quantum Electronics, Vilnius University, 
Sauletekio 9, building III, LT-2040 Vilnius, Lithuania}
\begin{abstract}

We exploit a slightly noncollinear second-harmonic cross-correlation scheme 
to map the 3D space-time intensity distribution of an unknown complex-shaped 
ultrashort optical pulse. We show the capability of the technique to 
reconstruct both the amplitude and the phase of the field through the coherence 
of the nonlinear interaction down to a resolution of 10 $\mu$m in space and 200 
fs in time. This implies that the concept of second-harmonic holography can be 
employed down to the sub-ps time scale, and used to discuss the features of the 
technique in terms of the reconstructed fields. 

\end{abstract}
\end{frontmatter}

\section{Introduction}

The study of ultrafast phenomena has been a major scientific priority
during last decades covering different topics such as the study of
radiation-matter interactions \cite{hullier}, transient response of 
molecules and atoms \cite{zewail}, coherent control of chemical reactions 
\cite{Baumert} or communication and information technology 
\cite{stegeman}. 
The growth of this field relies upon the development of sources of 
femtosecond radiation and of appropriate techniques able to provide 
time domain information in the femtosecond scale. 
However, during the interaction of 
short optical pulses with a nonlinear medium, different mechanisms
can lead to its reshaping into complex {\em spatio-temporal} structures
with non-trivial light
distribution \cite{PDT00}. As a
consequence, their complete characterization requires a method
capable of acquiring a snap-shot of their intensity distribution
in the whole 3-dimensional (3D; $x,y,t-z/c$) comoving frame. 

Most of the available methods for pulse diagnostic provide information of the WP
characteristics in a space of reduced dimensionality. The use of
frequency resolved autocorrelation techniques (i.e. FROG,
SPIDER) allows for the recovery of the temporal intensity and phase
profile of a given pulse but assumes uniform transverse spatial
distribution \cite{Trebino, Jaconis}. On the contrary, the
characterization of transversally localized beams
often relies upon the optical imaging onto CCD cameras, therefore
the temporal information is lost because of their integration times 
unavoidably larger than the optical pulse duration. 
Recently, a space-time characterization
method based on extended SPIDER technique has been developed
capable of resolving electric field characteristics in time and
along one spatial coordinate \cite{dorrer}. 
A quite direct way of obtaining spatio-temporal intensity profiles 
of a WP is to perform
measurements with a streak camera, which allows a
temporal resolution up to fractions of ps \cite{streak}. This
technique allowed the investigation of the dynamics of the breakup
along the pulse envelope of a large   elliptical beam propagating
into a saturable Kerr nonlinear medium \cite{Lantz02a,Lantz02b}.
However, also in this case, the space-time maps are intrinsically two 
dimensional (1 spatial + temporal dimensions).

A different approach to the problem considers the retrieval of the 
pulse shape through an all-optical processing by means of 
spatially resolved detection systems combined with gating techniques. 
The principle of the method is that of characterizing with spatial
resolution 
an optical field that is proportional to the product 
$E_O({\mathbf x},t)E_R({\mathbf x},t)$,
where $E_O({\mathbf x},t)$ is the object to be measured and
$E_R({\mathbf x},t)$ is a suitable reference pulse. Since the
product is different from zero only on the intersection of the
support of both fields\footnote{Rigorously, the support of a
function is the set of points where its value is different from
zero. For realistic optical fields that have exponentially
decaying tails we can define a ``practical'' support, defined as
the set of points in the space-time in which the field amplitude
is larger than $1/e$ times the peak value.}, by translating the 
reference with respect
to the object, we get the possibility of recording information
from different parts of the object. Among
the linear time gating techniques, light--in--flight holographic
recording has been the first technique which permitted the
recording of dynamically evolving light fields during propagation 
\cite{Denisyuk69,Abramson83,Abramson89}. Recently, this technique 
has been adapted to study the propagation of a 3 ps long pulse in
linear media \cite{Kubota02}. Linear probing techniques were
also exploited in order to obtain time resolved imaging, like the
probing of the birefringence properties of plasma by means of a
delayed, spatially-extended 100 fs pulses to investigate the
dynamics of laser pulse focusing in air \cite{Fujimoto99}. 

Nonlinear processes have been employed since long ago to resolve in
time the evolution of ultrafast phenomena. Among them, the quadratic
nonlinearity has been proved to be particularly versatile due to
the fact that it provides easily terms containing the product of
two optical fields. Recently, a type II degenerate parametric
amplification scheme has been employed to obtain time resolved 2D
images of a ps-pulse hitting a diffusing screen with 35 ps
resolution \cite{Devaux95}, thus yielding a 3D imaging. The same 
technique was later used to image an object embedded in a thick diffusing 
sample \cite{Devaux99}. 
Althought our setup is actually an improved version 
of that described in \cite{Devaux99}, we point out that our conceptual 
approach is different from the study of the propagation of a wave 
front. In fact, in our case the propagation variable is fixed. 

Our goal in this article is to demonstrate the potentiality of 
the optical gating technique to acquire a high resolution space-time 
map of short, focused WPs in their comoving reference frame. Furthermore, 
we devise the capability of the technique to reconstruct both the 
amplitude and the phase of the WP thanks to the coherence of the 
nonlinear interaction. 
We propose a method that is based on quadratic type I interaction in a
sum-frequency generation scheme either by a non-collinear second 
harmonic generation or by a collinear sum-frequency scheme. 
The latter has been used in \cite{jose}. 
Here we discuss the first option, showing that if the interaction
angle between the two interacting fields is small enough, then a
reliable space-time map of the object pulse can be obtained. 
This can be achieved if the duration of the gate is 
much smaller than that of the object to be imaged. 
A holographic interpretation of the method permits to gain insight 
into the process of up-conversion of the space-time slices of the 
object into the SF field, and to prove that the coherence of the 
SF process is able to reconstruct the wavefront in both amplitude 
and phase. Our results confirm this possibility. 
The theoretical discussion of the method is followed by section
\ref{experiment}, where we present the set-up and the
experimentally reconstructed space-time intensity profiles of a
parametric spatial soliton excited by a 1 ps light pulse. For our
setting, we estimate a mapping resolution of 200 fs in time and
about 10$\mu$m in space. 
The features of the technique are presented in section \ref{features},
pointing out the limitations that may arise and discussing the
possible implementations in each case. In the last section the
main conclusions are presented.

\section{Description of the technique: intensity and field reconstruction}
\label{crosscorrelation}

In this section we explicitly show how a non-collinear sum-frequency (SF) 
scheme can be exploited to get high resolution space-time intensity maps of an
unknown light wave packet with a space-time structure (object wave).
The recovery of a 3D intensity map is obtained by means of a short reference pulse 
which provides a time gating inside a nonlinear (NL) crystal, and generates a SF 
signal containing the information about a set of 2D slices of the object obtained by 
changing the reference delay. 
We first discuss the case for the reconstruction of the object intensity profile, 
and then we show how the intrinsic coherence of the SF process allows 
in fact for a truly holographic recording of the unknown object.

\subsection{3D Intensity profile mapping}

Let us denote the object ($\bar{E}_O$) and reference ($\bar{E}_R$) wave packets
as follows:
\begin{eqnarray}
\bar{E}_O&=& E_O(x,y,z,t)e^{i[\omega_1 t-k_z(\omega_1)z-k_x(\omega_1)x]}+c.c.\\
\bar{E}_R&=& E_R(x,y,z,t)e^{i[\omega_2 t-k_z(\omega_2)z+k_x(\omega_2)x]}+c.c.
\end{eqnarray} 
where the complex functions $E_O(x,y,t,z)$ and $E_R(x,y,t,z)$ are the slowly 
varying envelopes of two waves with frequencies $\omega_1$ and $\omega_2$. 
Note that in this form the equations describe two 
wavepackets propagating in the positive $z$ direction and colliding at
an angle $\theta = 2 \arctan ( k_x / k_z)$ in the $x-z$ plane (here 
$k = \sqrt{k_x^2 + k_z^2} = 2 \pi / \lambda_0$). 
For a SF generation process occurring inside a quadratic nonlinear crystal, 
the polarization source giving rise to the SF can be written as:
\begin{eqnarray}
P_{SF} \propto 2 E_O E _R e^{i [\omega_3 t - k_z(\omega_3) z ]} 
+ c.c.
\label{polarizzazione}
\end{eqnarray}
where the phase and energy matching conditions $k_z(\omega_3) = k_z(\omega_1) + 
k_z(\omega_2) $, $k_x(\omega_1) = - k_x(\omega_2)$ and $\omega_3 = \omega_1 + \omega_2$ 
have been used. 
The SF field propagates along $z$ direction and has a slowly varying envelope 
function that we will indicate by $E_{SF}(x,y,t,z)$. 
Now we introduce the following assumptions: 1) small depletion of both the $E_O$ and 
$E_R$ fields; 2) negligible diffraction and dispersion within the crystal; 3) equal 
group velocities of the object, reference and sum-frequency pulses, namely $u_O$, $u_R$ and 
$u_{SF}$, that is $u_O=u_R=u_{SF}=u$. 
Note that all these assumptions approximatively hold as long as the thickness 
of the crystal $\Delta z$ is small compared to the characteristic lenghts of the 
system (nonlinear length, dispersion and diffraction lengths, pulse walkoff 
length). 
These assumptions allow to find a travelling reference frame for all the propagating 
pulses by introducing the retarded time $\tau=t-z/u$. The general partial 
differential equations describing the interaction process then reduces to an ordinay 
differential equation for the envelope $E_{SF}$. If we also introduce a time delay 
$\tau_i$ on the reference wavefront, the equation takes the form: 
\begin{equation}
\frac{d E_{SF}(x,y,\tau,z)}{d z} = i2\sigma E_O(x,y,\tau,z)E_R(x,y,\tau-\tau_i,z)
\label{sumfrequency}
\end{equation}
where $\sigma$ is the nonlinear coupling term. 
The last equation is readily integrated and, if the mixing crystal is placed 
at position $z_0$, it reads: 
\begin{equation}
E_{SF}(x,y,\tau,z_0)=i\sigma 2 E_O(x,y,\tau,z_0)E_R(x,y,\tau-\tau_i,z_0)\Delta z
\label{campi}
\end{equation}
A deeper discussion about the meaning of this expression will be given in the 
following subsection. 
Here we just point out how the intensity profile of the SF field can be 
used to retrieve the 3D intensity map of the object. 
More precisely, for a given lag time $\tau_i$, the spatially dependent SF fluence 
profile (the CCD signal) $S(x,y,\tau_i,z_0)$ recorded just at the exit face of the 
mixing crystal is given by: 
\begin{equation}
S(x,y,\tau_i,z_0) \simeq (\sigma\Delta z)^2\int^{+\infty}_{-\infty}|E_O(x,y,\tau,z_0)|^2
|E_R(x,y,\tau-\tau_i,z_0)|^2 d\tau
\label{convolution}
\end{equation} 
This expression provides the convolution between the intensity profiles of the 
object and reference wave packets, lagged in time by $\tau_i$, and shows that the 
signal recorded by a CCD sensor is a {\it linear} function of the intensity of both 
the object and reference wavepackets. 
In the particular case in which the reference wave is spatially homogeneous in the transverse 
$x-y$ plane, and temporally much shorter than the object, we can write 
expression \ref{convolution} in the following form: 
\begin{equation}
S(x,y,\tau_i,z_0) \propto (\sigma\Delta z)^2 |E_R|^2 I_O(x,y,\tau_i,z_0)
\label{intensity}
\end{equation}
where $I_O = |E_O|^2$.

By imposing a set of delays ($\tau_i$, $i = 1...n$) to the reference WP with respect 
to the object, a reliable 3D reconstruction of the WP structure can be achieved 
by the collection of the $n$ images $S(x,y,\tau_i,z_0)$. By changing the plane 
$z_0$, the temporal evolution of the WP can also be obtained. 

Notice that the model described above does not take into account the dispersion of the 
mixing crystal. Therefore, the model as it is predicts no limits for the resolution 
of the map as long as arbitrary short reference pulses are available. 
Actually the real mixing crystal has a finite bandwidth that limits the spatiotemporal 
resolution of the obtainable maps. Therefore a careful evaluation of the dispersion 
characteristics of the mixing crystal have to be gauged as ultrashort pulses are 
either investigated or used as a reference. More details are discussed in section 
\ref{features}. 

\subsection{3D field reconstruction}

The method described above is not limited to the intensity reconstruction, but can be also 
implemented for the field reconstruction as can be stated from equation \ref{campi}. 
This point can also be explained in terms of a holographic description of the process, thus 
bringing to a deeper understanding of the imaging reconstruction process. 

As stated in \cite{Denisyuk99,Denisyuk00,staselko}, for a plane wave reference the 
generated wavefront through the NL interaction behaves 
as a conventional hologram recorded at frequency $\omega$ and illuminated with a 
radiation at $2\omega$. 
As a consequence, the position, scale, resolution and all the other properties of 
transformation of the reconstructed image can be predicted by means 
of the ordinary laws of holography. This hologram is recorded and reconstructed at a 
time, and exists only when light propagates inside the crystal. 
Yet we also point out that we obtain the 3D map of our pulse by collecting a set of 
independent 2D holograms by slicing the object pulse at different delays. 
Nevertheless, according to the holographic properties of the SF process, the
slicing can be done with the mixing crystal at any $z$ from the real object to be 
recovered and the information recorded is enough to reconstruct the slice. 

\begin{figure}
\epsfig{file=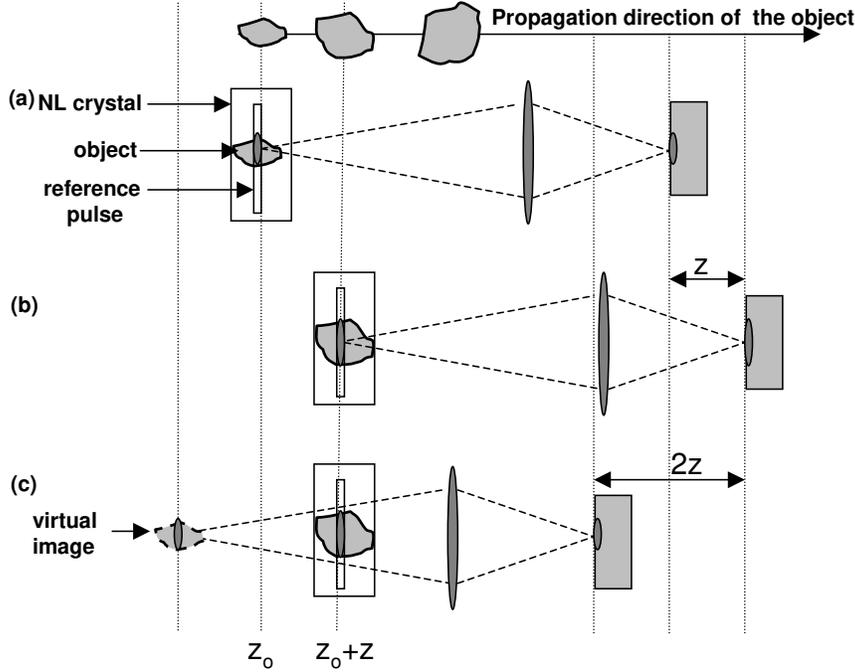, width=12cm, angle=0}
\caption{Three reconstruction schemes of a slice obtained from an object WP. 
In a) the slice is reconstructed as a real image (confocal configuration); 
in b) the slice is reconstructed after the object propagated a distance $z$; 
in c) the slice corresponding to the one imaged in b) is used to reconstruct 
the virtual image of the one imaged in a). }
\label{holoscheme}
\end{figure}

In order to clarify this point, let's assume to perform an experiment in which we
reconstruct a slice of an object WP (Fig \ref{holoscheme}.a). At any 
distance $z$ (Fig \ref{holoscheme}.b) the reconstructed WP will
correspond to the propagated version of the one at $z_0$.
Let us consider the case when the object WP has a bandwidth small enough with respect to the 
SF bandwidth that our simplified model applies, the pulse does not have appreciable 
angular dispersion\footnote{In the case of angular dispersion, the diffraction 
drives an effective group velocity dispersion also in the vacuum, then the spatial 
evolution cannot be separated from the temporal one (see for example \cite{sonajalg}, 
\cite{zozyula})} and the spatial and the temporal evolution can be separated. 
Under these assumptions, any slice recorded in 
the confocal configuration (see Fig. \ref{holoscheme}.a) 
could also be reconstructed when the mixing crystal is displaced 
at position $z$ and the field to be converted is the propagated one. 
This is possible by exploiting the holographic properties, provided that the detecting 
system is set to reconstruct the virtual image of the slice. This image is placed at a 
distance 2$z$ far from the mixing crystal (see Fig. \ref{holoscheme}.c), its 
transverse size being identical to that of the object (see \cite{Denisyuk00}, 
and also note that we are working in the degenerate case, $\omega_1 = \omega_2$). 

The possibility to recover the intensity profile of a virtual image comes from the 
complete wavefront ({\it field}) reconstruction arising from the SF process coherence 
and contained in Eq. \ref{sumfrequency})\footnote{We suggest for example the 
possibility to use an interferometric or heterodyne device in order to obtain a 
complete charaterization of the field. By measuring the field instead of the 
intensity, a remarkable increment of the dynamical range is also possible.}. 

This important property also allows to get intensity maps with a
wider dynamical range. For example, in the cases when the object is so 
intense so that the undepleted pump approximation does not hold, we can get the
WP profile by displacing the NL crystal to a plane where the intensity
is reduced and then recover the intensity profile at the desired
plane $z_0$ by suitably moving the imaging system (Fig \ref{holoscheme}.c)).

\section{Experimental results}
\label{experiment}

\begin{figure}
\centering
\epsfig{file=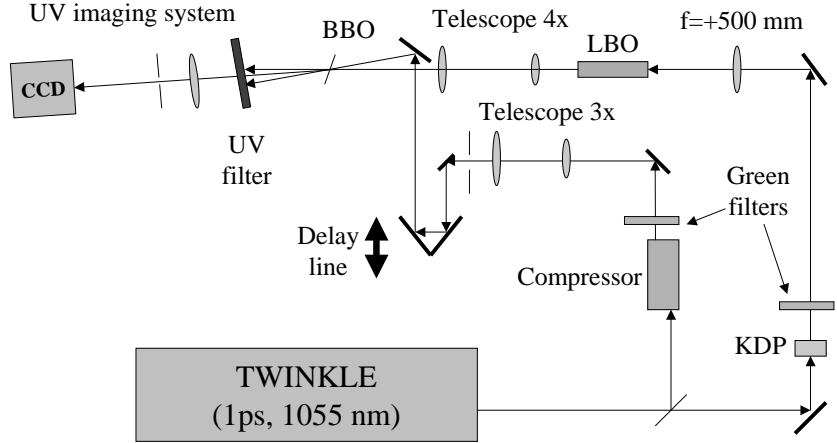, width=14cm}
\caption{\label{STimaging} A sketch of the experimental set-up used 
to retrieve the space--time intensity profiles.}
\end{figure}

We prepared several experiments in order to prove the possibility to recover 3D maps
of short WPs and to show their holographic properties.The experimental set-up is 
sketched in Fig. \ref{STimaging}. 

The 1 ps pulses of 
a Nd:glass laser source (TWINKLE, Light Conversion, wavelength 1055 nm) are splitted on 
two lines by means of a beam splitter. 
In the first line, the laser pulse is frequency doubled in a KDP crystal 
and focused on a 15mm-long litium triborate (LBO) crystal. 
For pulses about 1 $\mu$J in energy , spatial solitons are formed in 
this crystal because of the optical parametric generation process \cite{DiTrapani98}.
The gaussian temporal profile causes the spatial soliton to be formed only in the central 
part of the pulses where the intensity is higher \cite{Barthelemy02}. 
This leads to a non-trivial space-time structure in the output pump wave packet 
\cite{Minardi03}, which we will consider as the object. 
Furthermore, the object has been magnified 4 times by means of a two-lens 
telescope imaging the LBO exit face into the NL mixing crystal ( a 100 $\mu$m thick, 
$\beta$ barium borate crystal, BBO). The beam expansion 
has been necessary to: {\it i}) reduce the beam intensity and therefore fulfill 
the undepleted mixing requirement; {\it ii}) to avoid information loss related to 
the finite spatial resolution of the mixing process. 
In the second line, a 200-fs reference pulse is produced at the wavelength 
of 527.5 nm by means of a second-harmonic pulse compressor \cite{Stabinis91} 
and expanded 3 times by means of a telescope. 
Both lines are recombined in the thin BBO crystal, cut and oriented to generate the 
non-collinear SF from the object and reference beams.
An external incidence angle of $\sim6.5^\circ$ between the propagation 
directions has been chosen. 
The delay between the two pulses can be varied by means of a suitable delay line 
placed on the reference pulse line. 
The non-collinear SF radiation is spatially selected by an aperture, then selected 
in frequency by means of coloured filters 
and the plane of the BBO is imaged onto the CCD sensor (PULNIX TM6CN). 
Finally, both the BBO crystal and the imaging system could move independently on 
a rail along the optical axis.

At first we focused the object inside the BBO crystal and adjusted the position 
of the imaging system as in Fig. \ref{holoscheme}.a). 
By recording the spatial profile of the SF radiation as a function of the 
reference pulse time lag (steps of $\Delta \tau_i = 66.6$ fs has been used, where 
$\Delta \tau_i = \tau_i - \tau_{i-1}$), we have retrieved the 3D isointensity 
maps of the object pulse at the exit face of the LBO crystal. 

\begin{figure}
\centering
\mbox{\subfigure[]{\epsfig{file=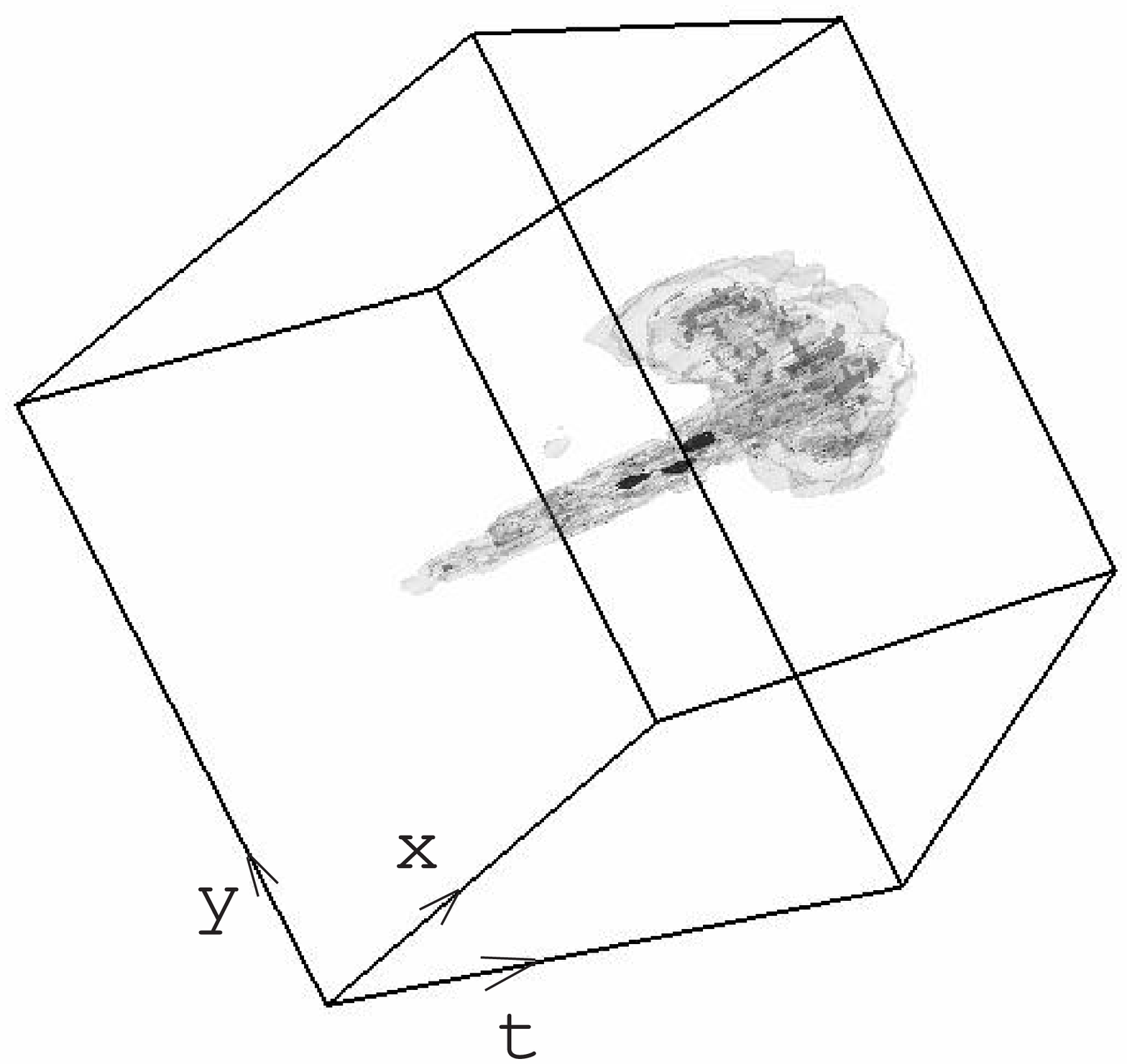, width=5cm}}\quad
\subfigure[]{\epsfig{file=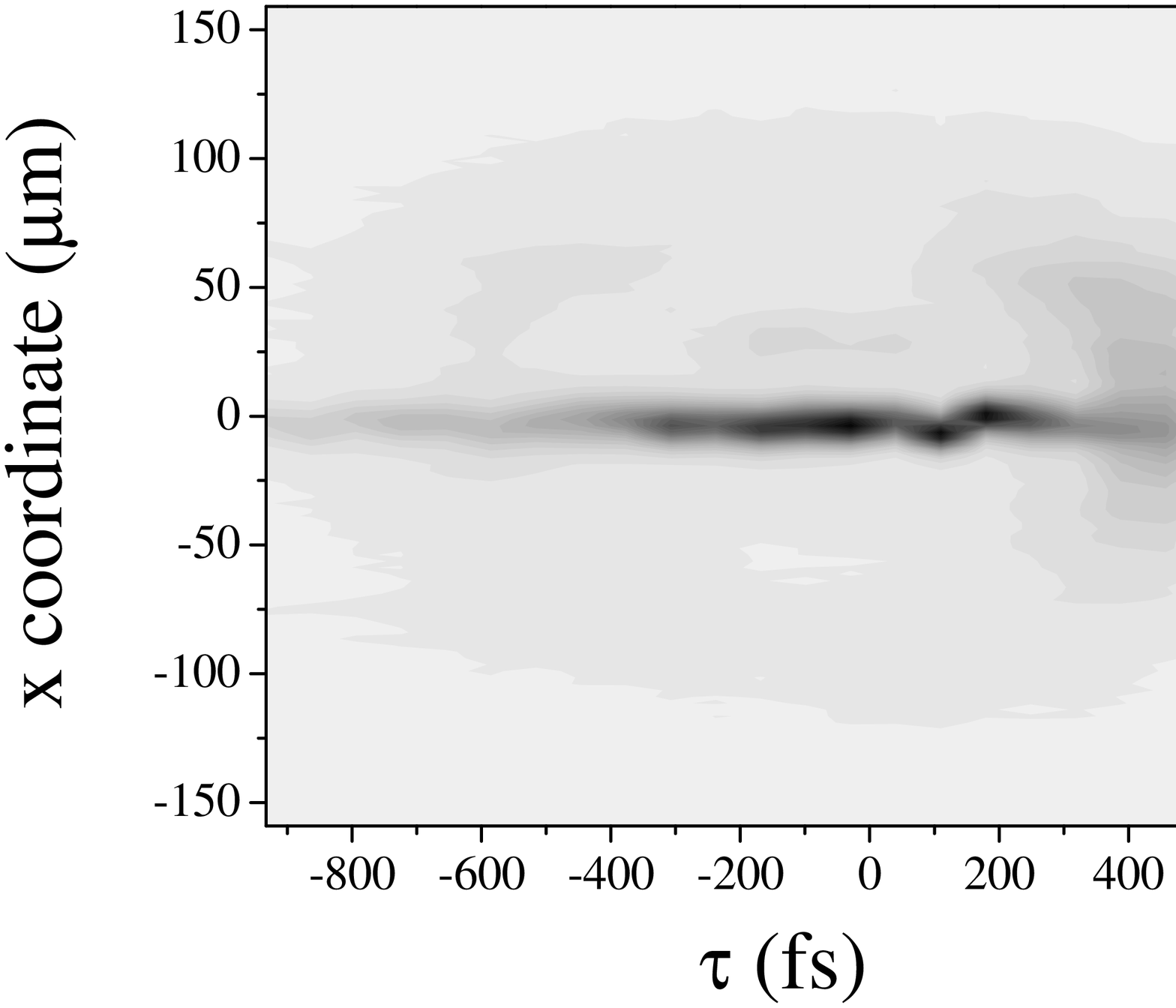, width=6.5cm}}}
\caption{\label{isointensity} a) Isointensity map of the object pulse 
as reconstructed from experimental data. 
The rendering has been performed with the Linux-based program GOPENMOL. 
b) A contour map of the $(x,\tau)$ section of 
the same pulse.}
\end{figure}

Fig. \ref{isointensity}.a shows three different intensity levels of the pulse in the 
$(x,y,\tau)$ space, while the corresponding contour plot of the $(x,\tau)$ 
plane section is depicted in Fig. \ref{isointensity}.b. 
The space-time maps clearly show that the pulse is formed by a spatially focused 
structure followed by a diffracted tail.  
The accuracy of this reconstruction has been successfully tested by comparing 
the experimental plots with the results of a 3D-numerical simulation 
of the formation process of the object pulse (see \cite{Minardi03}). 

\begin{figure}
\centering
\epsfig{file=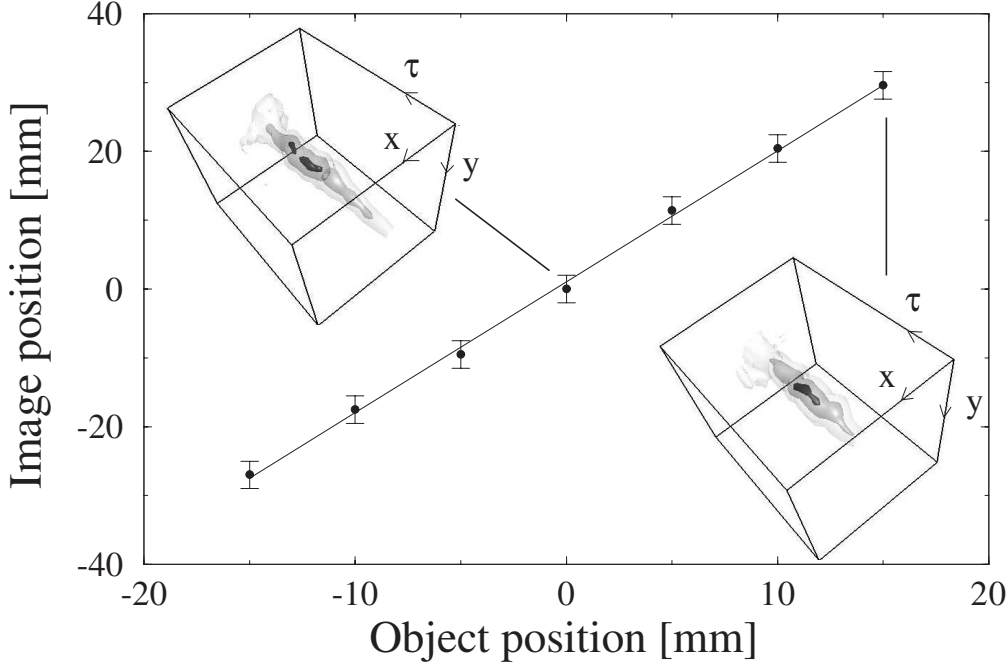, width=9cm, angle=-90}
\caption{\label{holoprop} The positions of the virtual images reconstructed as 
described in Fig. \ref{holoscheme} c) for several positions of the real object 
with respect to the BBO crystal. 
The two inserts show the real image ($z = 0$, above left) and the virtual image 
as reconstructed after the maximum propagation distance (15 mm, bottom right). 
All the distances are measured with respect to the position of the BBO crystal 
(positive values, image plane positions beyond the crystal in the direction of 
the beam propagation). The points fit a straight line with a slope of about 2, 
as expected from holography.}
\end{figure}

In order to check for the validity of the holographic interpretation 
in this process, we have reconstructed the same WP of Figure 
\ref{isointensity} by moving the NL crystal a distance z far from the previous 
position and by scanning the position of the imaging system (CCD+lens) along 
the SF propagation direction in order to find out the position of the (virtual) 
reconstructed image. First we fixed the time lag between the object and the reference 
and selected a slice of the object corresponding to a narrow focused spot 
of $\sim 40 \mu$m in diameter inside the BBO crystal (corresponding to case a) 
in Fig. \ref{holoscheme}). 
Furthermore, the displacements $z$ of the BBO crystal have been chosen large enough 
to ensure that the propagated wavefronts had lost the transverse structure. 
As a rough estimate, for the focused part of the object, the Rayleigh range is about 
7 mm long, while we spanned distances from -15 mm to 15 mm. 
In Fig. \ref{holoprop}, data show the position of the virtual image against the 
position of the real object from the crystal (see Fig. 1.c)), showing a remarkable 
fit to a straight line which slope is close to the value of 2, according to what 
discussed in the precedent section. 
The error bars shown in the figure indicate the estimated uncertainties 
in the reconstructed image plane position, measured by scanning the 
whole Rayleigh range and by extracting the position of the central point. 
In the same figure the two insets 
show two isointensity maps recovered for the case when the imaging
system is focusing directly the BBO plane (z=0) and when the BBO
crystal has been moved a distance z=15 mm (the maximum propagation
distance we imposed to the BBO). 
The agreement between the two intensity profiles proves the reliability of our 
method to work with virtual images in recovering 2D slices of WPs like those we used here. 

\begin{figure}
\centering
\epsfig{file=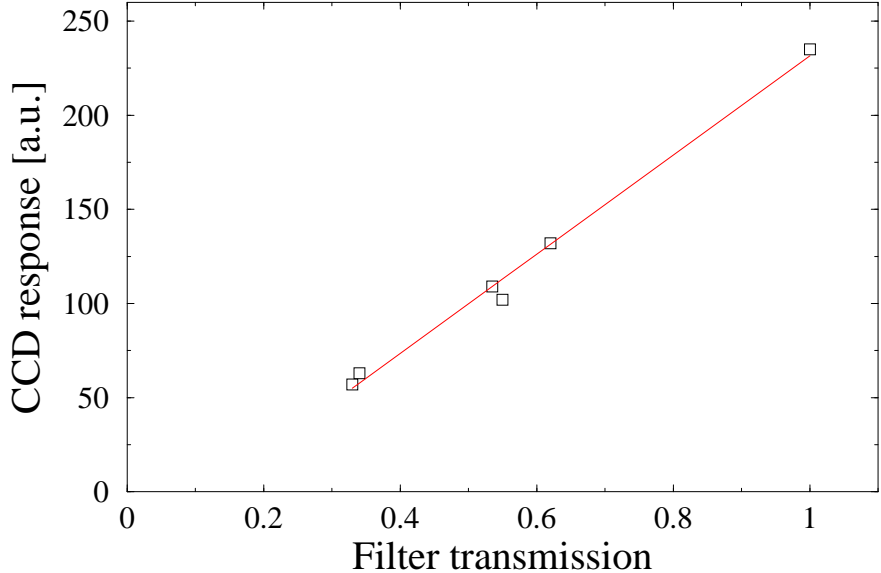, width=8cm, angle=-90}
\caption{\label{linearity}: The intensity measured by the CCD (arbitrary units) 
plotted against the transmission of the filters used to reduce the object intensity. 
The linear behaviour of Eq. \ref{convolution} confirms that the 
measurements were performed in the undepleted--field regime. }
\end{figure}

In order to verify that we have always been operating in the 
undepleted regime, we have measured the dependence of the generated SF 
field as a function of the object-pulse energy at $z = 0$. 
The results are presented in Fig. \ref{linearity}, where the peak fluence (as 
registered from the CCD images) is plotted vs the transmission of the neutral-density 
filters that attenuate the object. The costant slope confirms the 
absence of any saturation in the SF process. Data also indicate that a slight 
overestimate of the background has been done during the profile acquisition 
(zero SF field is found for 15 \% filter transmission).

\section{Features of the technique}
\label{features}

In the previous sections we have proved that the described technique is able to perform 
the 3D mapping of objects similar to those we used for our measurements. 
Anyway, one can expect that in general the reliability of the technique to recover the  slices could 
be affected by particular effects dealing with the interaction geometry, the 
mixing crystal, and the features of the interacting pulses, in particular for 
broad-band, ultra-short and chirped pulses. 
Therefore we discuss here three main features of our technique, namely, the duplication 
bandwidth of the mixing crystal, the interaction angle and 
the reference pulse shape, that bring to devise some possible limitations 
to the fidelity of the technique.

\subsection{Space-time resolution}

As we briefly mentioned above, the interaction model we describe  does not take into account 
the actual finite bandwidth of the mixing crystal, limiting the spatiotemporal 
resolution of the maps. 
In fact, in our experiment the BBO crystal has been chosen thin enough that 
the converted bandwidth was large compared to the object one. 

As a matter of fact, the spatio--temporal resolution of the slices is 
dictated by the maximum range of angles and frequencies over which the 
conversion is effective (for very broad--band objects the angular blurring 
that could arise from the phase matching condition should be considered). 
The key parameter here is the maximum phase mismatch between the fundamental and 
the SF waves at which the conversion efficiency vanishes, $\Delta k_{max}$. 
On the basis of the existing literature \cite{niko}, we have estimated the 
maximum bandwidth converted by the crystal as the FWHM of the efficiency 
curve, yielding to a temporal bandwidth of about $630$ cm$^{-1}$ and an 
angular bandwidth of about $510$ cm$^{-1}$. 
By considering only the angles for which the conversion efficiency is 
higher than 1/e of the maximum value \cite{devaux}, we obtain the remarkable 
resolution of details approximately 10 $\mu$m in size (100 lines/mm). 
We can notice that the holographic interpretation of the wavefront 
reconstruction leads to an easy description of the imaging process.
Furthermore, our optical system could be used in principle to achieve 
a microscopy of ultrashort objects shorter than the one we used here \cite{jose}, 
although the limit of very shot WP is detrimental for the holographic 
reconstruction of the phases.

\subsection{Interaction angle}

A point to be discussed concerns the influence of the noncollinear geometry and 
pulse chirp on the reconstruction of the hologram. In 
principle the holographic interpretation is strictly valid for collinear geometry 
only (when $k_2 = 2 k_1$) \cite{Denisyuk00}. 
Although this condition is not strictly fulfilled in our experiment, we have 
maintained the interaction angle small enough to make this disturbance negligible. 
However, a temporal chirp in the object pulse could give rise to a spatial 
phase distortion of the SF wavefront \cite{Stabinis91b}. 
Although this does not affect the results obtained with the confocal configuration 
(see Fig. \ref{holoscheme}.a), it could distort the maps obtained in the holographic 
one (Fig. \ref{holoscheme}.c). We expect that this effect is really 
rampant only when complicated ultrashort pulses with strong chirp are considered, or 
large interaction angle are employed. 
We are confident that our system was operated far from this limit, since, 
as we checked from the data, our experimental holograhic maps do not show 
any appreciable enhancement of the astigmatism. 

\begin{figure}
\centering
\epsfig{file=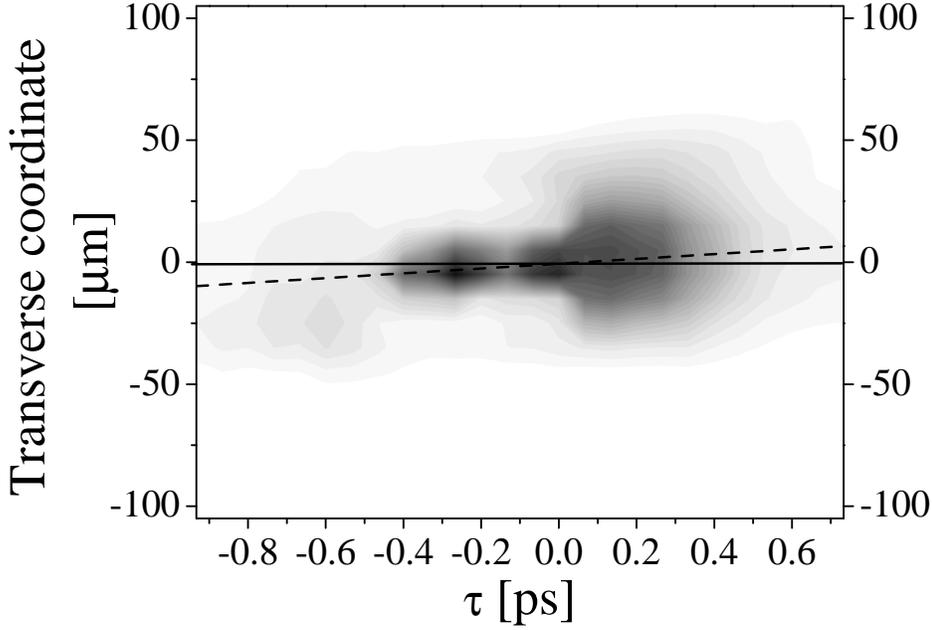, width=14cm}
\caption{\label{skewed} Space time map obtained with an external angle of 16 deg 
between the reference and object pulse directions of propagation. The dashed line 
represents the map skewing as calculated from equation \ref{angoli}.}
\end{figure}

As it is well known, the non-collinear SF scheme is largely used to get single-shot 
autocorrelation traces of ultrashort pulses \cite{Gyuzalian79}.
In fact, because of the geometry of the interaction, it is easy to show 
that the time of the interaction between the object and reference pulses 
depends on the transverse coordinate of the intersecting planes (see 
\cite[pag. 426-428]{zewail}).
Therefore, if the interacting angle is large, we expect that distortions of the 
spatial profile along this direction could affect the space-time maps. 
However, it turns out that in the confocal configuration (\textit{i.e.} 
as in Fig. \ref{holoscheme}.a) the effect of the interaction angle 
is merely that of skewing the intensity map, so that the actual time-axis 
of the object pulse is not parallel to that measured on the delay line. 
This is evident in the map depicted in Fig. \ref{skewed}, 
obtained with an external crossing angle of $\sim 16^\circ$.

In the paraxial approximation, the slope of the object-pulse 
time-axis on the space-time map is given by:
\begin{equation}
\tan\beta=\frac{c}{n}\sin\frac{\gamma}{2}
\label{angoli}
\end{equation} 
where $\gamma$ is the angle between the propagation directions of the 
object and the reference wavepackets inside the crystal, and $n$ is the 
refractive index of the medium. It is easy to recognize this formula 
as the calibration relation of the non-collinear, single-shot intensity 
autocorrelator \cite{Gyuzalian79,zewail}.

\subsection{Reference pulse shape}

Finally the last point to be discussed concerns the shape of the reference pulse. 
In our experiments it was only slightly affected by the presence 
of satellites because of the generation {\it via} the pulse compressor 
\cite{Stabinis91}, the amplitude of these structures being small enough to be 
considered negligible to our aim. 
Actually, the recovered intensity profile of the object would be distorted in case 
the reference pulse has a structure more complicated than the single
peak envelope, as Eq. \ref{convolution} clearly points out. 
This suggests that a careful measurement of the reference shape 
(for example carried out by means of an autocorrelation technique) can be used 
to retrieve the real intensity profile by means 
of a deconvolution procedure. We point out, however, that any deconvolution 
unavoidably introduces an extra spurious noise which could degrade the final 
quality of the mapping \cite{deconvolution}).

\section{Conclusions}

We have shown that the 3D intensity maps of optical WPs with a complex 
space-time structure can be retrived by an optical gating technique. 
The method allows the reconstruction of the WP in its comoving reference 
frame and, by exploiting the holographic properties of a slightly non-collinear 
degenerate, sum-frequency process, we have shown that a complete amplitude 
and phase reconstruction is actually obtained. 
The maps are obtained by suitably imaging the second harmonic radiation 
obtained by cross-correlating an object pulse with a much shorter plane 
wave-packet delayed in time. 
The holographic properties of the generated radiation have been carefully tested 
experimentally, and the distorsions of the intensity maps introduced by the 
non-collinear geometry have been discussed in detail. 
Finally, theoretical considerations point out that the ultimate 
resolution of our set-up is in the order of 100 lines/mm in space, and of a 
few fs in time. However, the choice of the reference pulse limits 
the actual time resolution to about 200 fs.

We forsee that the developed technique will benefit all the fields where 
a space-time mapping of light pulses is relevant, such as the investigations
on the reshaping of ultrashort pulses propagating in non-linear materials
\cite{Lantz02a,Lantz02b,PDT00, Minardi03, jose}. 
Moreover, the holographic features of the technique might be exploited
to fully reconstruct the field of an object pulse and its evolution during the propagation. 

This work was partially supported by MIUR (COFIN01 and FIRB01), the European 
Commission EC-CEBIOLA project (ICA1-CT-2000-70027) and DGI BFM2002-04369-C04-03 (Spain).  
The work of J.T. is supported by a postdoctoral grant from Ministerio de Educacion Cultura y Deporte (Spain). 

$\dagger$ Permanent address, Department of Physics and Nuclear Engineering, Universitat Polit\'ecnica Catalunya, 08222 Terrassa, Spain.

\end{document}